\documentclass[%
reprint,
nofootinbib,
 amsmath,amssymb,
]{revtex4-1}

\usepackage[utf8]{inputenc}
\usepackage[T1]{fontenc}
\usepackage{ulem}
\usepackage{graphicx}% Include figure files
\usepackage{bm}% bold math
\usepackage{physics}
\usepackage{natbib}
\usepackage[colorlinks=true,linkcolor=blue,citecolor=blue,urlcolor=blue]{hyperref}

\begin{document}

\title{Diffusion of excitation and power-law localization in long-range-coupled strongly
  disordered systems} 

\author{K. Kawa}
\email{Karol.Kawa@pwr.edu.pl}
\author{P. Machnikowski}
\email{Pawel.Machnikowski@pwr.edu.pl}
\affiliation{Department of Theoretical Physics, Wrocław University of Science and Technology, 50-370 Wrocław}

\begin{abstract}
% \begin{description}
% \item[Keywords]
We investigate diffusion of excitation in one- and two-dimensional lattices with random
on-site energies and deterministic long-range couplings (hopping) inversely proportional
to the distance. Three regimes of diffusion
are observed in strongly disordered systems: ballistic motion at short time, standard
diffusion for intermediate times, and a stationary phase (saturation) at long times. We
propose an analytical solution valid in the strong-coupling regime which explains the
observed dynamics and relates the ballistic velocity, diffusion coefficient, and asymptotic
diffusion range to the system size and disorder strength via simple formulas. We show also that
in the long-time asymptotic limit of diffusion from a
single site the occupations form a heavy-tailed power-law distribution.
% \end{description}
\end{abstract}
%
% \keywords{Suggested keywords}
%                             
\maketitle

\section{Introduction}
\label{sec:introduction}
The seminal paper of Anderson \cite{AndersonP.1958} anticipated lack of diffusion in
particular random lattices in three dimensions (3D), starting the topic of Anderson
localization of (quasi)particles. % which remains still compelling.
Although the original discussion concerned systems with power-law couplings, much of the
subsequent research dealt with tight-binding-like models with nearest-neighbor coupling
and on-site disorder.
Within that framework, Mott and Twose \cite{Mott1961} proved the lack of diffusion in one dimension (1D).
Later, Abrahams~{\em et~al.} \cite{Abrahams1979} proposed single parameter scaling
hypothesis and proved the absence of diffusion in two dimensions (2D), confirming at the
same time localization in one dimension and localization-delocalization phase transition
in three dimensions with respect to the system size. 
Nonetheless, the single parameter scaling hypothesis remains an approximate result,
demands finite range hopping, uncorrelated disorder and time-reversal symmetry.  
It still leaves space for the possibility of diffusion in specific models, even for dimension less or equal to two.

Beside the tight-binding model with on-site disorder and nearest-neighbor inter-site
coupling, other theoretical models were studied from the point of view of localization and
metal-insulator transition \cite{Rodriguez2003,Roati2008}. 
Among others, much interest was devoted to models of uncorrelated diagonal disorder with
long range hopping of the power-law character $\propto 1/r^\mu$
\cite{Rodriguez2003,Dominguez-Adame2004}. 
Such a model represents several physical systems.
For instance, Levitov \cite{Levitov1989} analyzed delocalization of vibrational modes in a
3D crystal with dipole interaction $\propto 1/r^{3}$. 
Subsequently, the energy transfer in several systems has been explored extensively within similar models.
In biological light-harvesting systems, energy transport to the reaction center is
mediated by dipole interactions $\propto 1/r^3$
\cite{Celardo2012,Sarovar2010,Mohseni2008}. 
Furthermore, many-body systems of nuclear spins with the same $1/r^3$ interaction have been
studied in the context of localization \cite{Alvarez2015}. 
A model with both the the short-range hopping between adjacent sites and long-range
dipole-dipole coupling was used for describing energy transfer in self-assembled nano-rings
\cite{Somoza2017}. 
The long-range coupling was shown to stabilize the system against disorder.
Moreover, energy transfer has been observed experimentally in the planar quantum dot (QD)
ensembles \cite{DeSales2004}. 
It has been found that the transfer is even better for low-density samples i.e. of greater
inter-dot distances preventing carriers from tunneling between QDs. 
Thus, a mechanism different from quantum tunneling must be proposed for explaining the
energy transfer,  
and a plausible explanation seems to be the long range dipole-dipole coupling.
In fact, an ensemble of QDs seems to be a coupled system, as its radiation properties
cannot be explained as a sum of single emitters \cite{Scheibner2007,Kozub2012}.  
The fundamental coupling between the emitters emerges here from their interaction with a
common electromagnetic reservoir, leading to the dispersion-force coupling $\propto
\cos(kr)/(kr)+\sin(kr)/(kr)^2+\cos(kr)/(kr)^3$
\cite{stephen64,lehmberg70a,Varfolomeev:1971tm}, where $k$ is the resonant wave number,
which reduces to the usual $1/r^{3}$ 
dipole interaction on short distances but is dominated by the $1/r$ term at distances
larger than the wave length resonant with the optical transition.

A renormalization group analysis by Rodr\'{\i}guez \emph{et al.} \cite{Rodriguez2003},
proves the existence of extended states in the model of uncorrelated diagonal disorder and
power-law hopping $\propto 1/r^\mu$, with an exponent $\mu$ greater than the dimension of
the system. 
Extended states appear in such a system in the thermodynamic limit in the vicinity of the
energy band edge, even in 1D and 2D. 
Such a model was also used for investigating the wave packet propagation in 1D
\cite{DeBrito2004}. 
Time evolution of the wave packet is described here by its mean square displacement
and the participation ratio.  
The localization-delocalization transition occurs as a function of the disorder magnitude for $1<\mu<3/2$.
For $\mu>3/2$ wave packet tends to localize because of the short range of hopping
decreasing while $\mu$ increases. 
The particular case of $\mu=1$, relevant to the dispersion forces at large distances,
shows interesting properties already in the absence of disorder, showing a combination of
diffusive and super-diffusive transport \cite{Kloss2019}. However, in disordered systems this case
has been studied only marginally.  It is known that it shows certain criticality features,
like the divergence of the critical disorder strength needed to localize the upper band
edge \cite{DeMoura2005a}.  

In this paper we consider the dynamics of a single excitation in arrays of two-level
systems with the long-range $1/r$ hopping and strong diagonal disorder. We show that a
system of finite size in this limit shows three consecutive phases of excitation transport: a ballistic
one, followed by normal diffusion, and finally saturation of the mean-square diffusion
range. We analyze also the average distribution of the excitation in the asymptotic
(saturation) phase and demonstrate a heavy-tail power law quasi-localization around the
initially excited site.
We point out that in the strong disorder limit the excitation transfer is dominated
by the direct coupling with the initial site, which allows us to reduce the model to a
simplified form, which is exactly solvable in a certain range of parameters. In this way
we are able to relate the transport parameters (ballistic speed, diffusion coefficient,
and asymptotic diffusion range) to the system size and disorder strength. 

The organization of the paper is as follows.
In Sec.~\ref{sec:model}, we describe the physical system and theoretical model.
In Sec.~\ref{sec:simulation_results} we present the results of our numerical simulations. 
The approximate analytical solution for strongly disordered systems is presented in
Sec.~\ref{sec:solution}, analyzed in Sec.~\ref{sec:regimes} and discussed in
Sec.~\ref{sec:discussion}. Final conclusions are presented in Sec.~\ref{sec:conclusions}.

\section{The System and the Model}\label{sec:model}

\begin{figure}[tb]
\begin{center}
    \includegraphics[width=0.8\columnwidth]{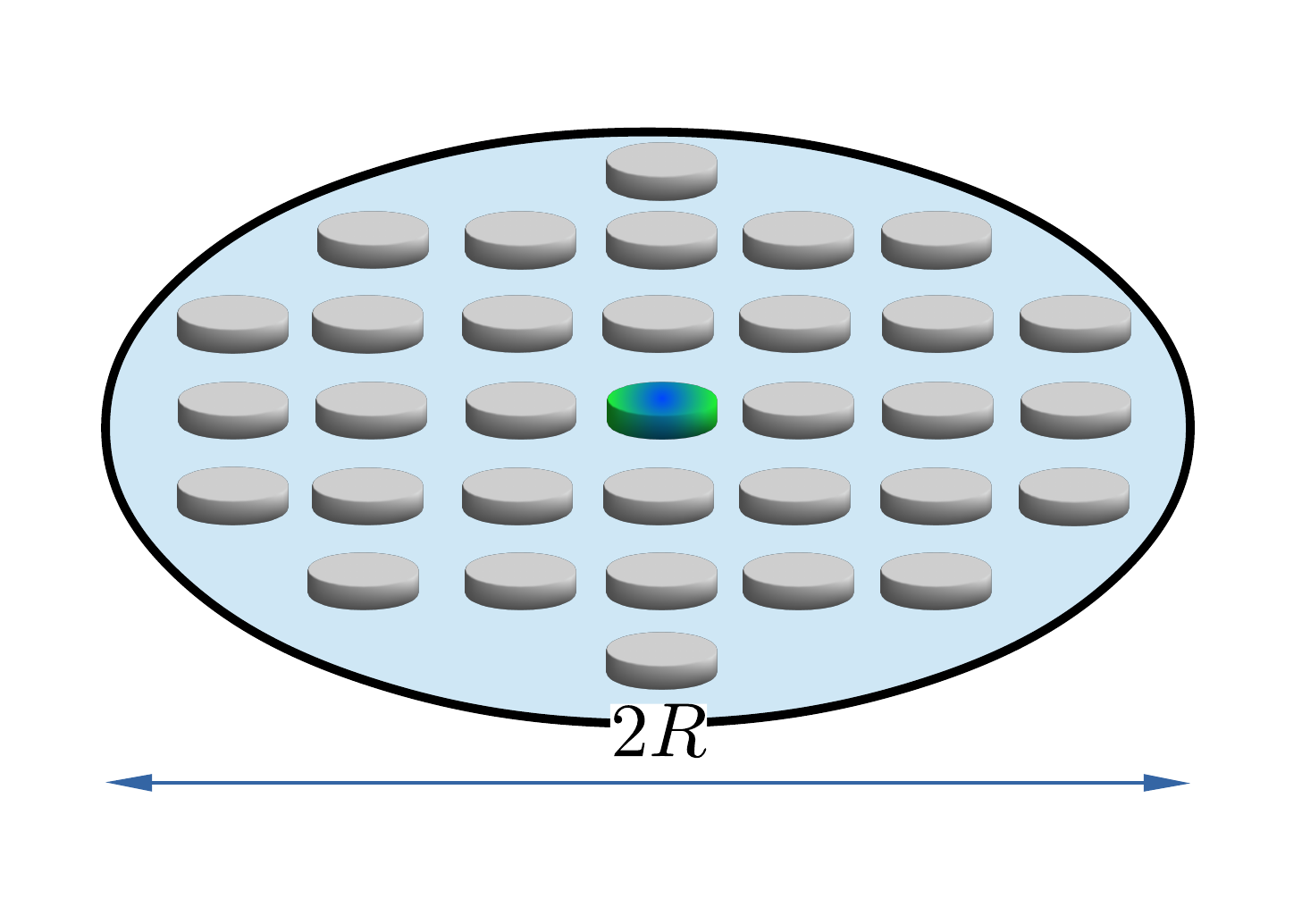}
\end{center}
    \caption{The system geometry for $d=2$: sites forming a regular lattice on a circular
      mesa. The initially excited site is marked by colors.} 
    \label{fig:system}
\end{figure}

In this section, we describe the physical system under study and introduce a theoretical
model used for numerical simulation.

We study a $d$-dimensional ($d=1,2$) system of $N$ sites on a regular lattice. The system occupies a
disk of radius $R$ for $d=2$ and a line segment of length $2R$ for $d=1$ (Fig.~\ref{fig:system}
shows the geometry for $d=2$).
The number of sites is related to the dimensionless radius of the system (in units of the
lattice constant) by $N = \zeta_d R^d$, where $\zeta_d$ is a number that depends on the
space dimensionality $d$. 

The system is described by the Hamiltonian
\begin{equation}
    H = J\left(\sum_\alpha \epsilon_\alpha \dyad{\alpha}{\alpha} + \sum_{\alpha \beta}
      V_{\alpha\beta} \dyad{\alpha}{\beta}\right), 
    \label{eq:hamilt}
\end{equation}
where $\ket{\alpha}$ represents a basis state localized at the site $\alpha$, $J$ sets the
overall energy scale, $J\epsilon_\alpha$ is the corresponding on-site energy, and
$JV_{\alpha\beta}$ is the coupling between the site states $\alpha$ and $\beta$. 
The dimensionless energies $\epsilon_\alpha$ (in units of $J$) are uncorrelated normally
distributed random variables of zero expected value and standard deviation $\sigma$. 
The coupling $V_{\alpha\beta}$ has a long-range character,
\begin{equation}
V_{\alpha\beta} = \left\{\begin{array}{ll}
    \dfrac{1}{|\mathbf{r}_\alpha - \mathbf{r}_\beta|}, & \textrm{for } \alpha\neq\beta, \\
    0, & \textrm{for }\alpha=\beta,
\end{array}\right.
\end{equation}
where $\mathbf{r}_\alpha$ is the dimensionless position of the site $\alpha$ (in units of the
lattice constant).

The central site is initially excited. 
The diffusion of the excitation is described by its mean square displacement (MSD) from the origin of the system
\begin{equation}
    \left\langle r^2(t) \right\rangle = 
\left\langle \sum_\alpha |c_\alpha(t)|^2 r^2_{\alpha}\right\rangle,\label{eq:r2_aver}
\end{equation}
where $\langle \dots \rangle$ denotes the average over disorder realizations, 
$c_{\alpha}(t)$ are the coefficients of expansion of the system state in the localized basis, 
\begin{equation}\label{Psi-expansion}
\ket{\Psi(t)} = \sum_\alpha c_\alpha(t)\ket{\alpha},
\end{equation}
and the central
site corresponds to the $\mathbf{r}_0 = \mathbf{0}$.

\section{Simulation results}\label{sec:simulation_results}

In this section we present the results of numerical simulations for the model described in
Sect.~\ref{sec:model} in one and two dimensions. The system evolution is found by exact
numerical diagonalization of the system Hamiltonian. 
We investigate the character of the excitation diffusion as a function of the system
parameters, i.e., the number of sites and the magnitude of the disorder
characterized by its standard deviation~$\sigma$. 

\begin{figure}[tb]
\begin{center}
    \includegraphics[width=\columnwidth]{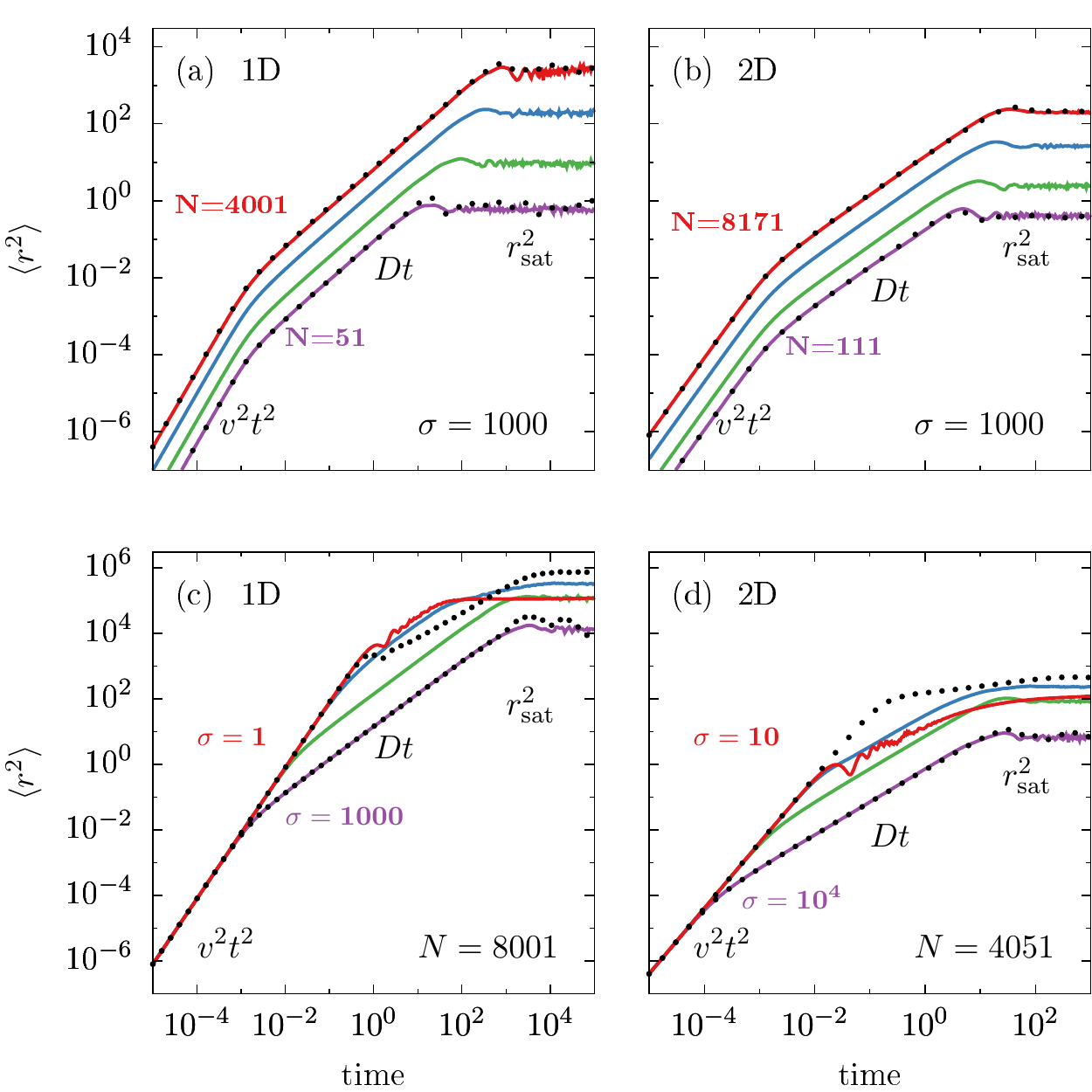}
\end{center}
    \caption{The MSD of the excitation from the central site as a function
      of time: (a,b)  for different sizes of the system but the same disorder $\sigma=1000$;
    (c,d) for different disorder magnitudes but the same system sizes, as shown. Panels
    (a,c) and (b,d) show the results for $d=1$ and $d=2$, respectively. Dotted lines show
    the results from the ``central atom model'' (Sec.~\ref{sec:analytic}). All the results
    are averaged over 1000 repetitions.}
   \label{fig:average_square_distance}
\end{figure}
The MSD of the excitation from the central site is shown in
Fig.~\ref{fig:average_square_distance} for different system sizes and disorder strengths. 
Three consecutive regimes of the excitation transport can be seen.
First, the transport is ballistic with a certain velocity $v$,
\begin{equation}
    \langle r^2 (t)\rangle = v^2t^2, \quad \textrm{for}\quad 0<t<t_0.
    \label{eq:ballistic}
\end{equation}
The velocity depends on the system size, as can be seen in
Fig.~\ref{fig:average_square_distance}(a,b), but not on the disorder
[Fig.~\ref{fig:average_square_distance}(c,d)].
At a certain cross-over time $t_{0}$, the normal diffusion with the diffusion coefficient $D$ sets on,
\begin{equation}
    \langle r^2 (t) \rangle = Dt, \quad \textrm{for}\quad t_0<t<t_1.
    \label{eq:diffusive}
\end{equation} 
As demonstrated by our simulation results in Fig.~\ref{fig:average_square_distance}, the
diffusion coefficient $D$ grows with the system size and decreases as the disorder
amplitude grows. 
From continuity requirement at the crossover one gets 
\begin{equation}
t_{0} = D/v^{2}.
\label{eq:t0}
\end{equation}
Simulations show that this value is size-independent but increases with the disorder strength.
Finally saturation is reached at the second cross-over time $t_{1}$.
\begin{equation}
    \langle r^2 (t) \rangle = r^2_{\mathrm{sat}}, \quad \textrm{for}\quad t>t_1.
    \label{eq:saturation}
\end{equation}
Obviously, the diffusion range must be limited in a finite system. However, the values
of $r^2_{\mathrm{sat}}$ presented in Fig.~\ref{fig:average_square_distance}, although
increasing with the number of nodes, are always considerably lower than the system
size and decrease as the disorder grows. Again, the cross-over time is fixed by
continuity, 
\begin{equation}\label{eq:t1}
t_{1}=r^2_{\mathrm{sat}}/D.
\end{equation}

In order to study the dependence of the dynamical characteristics $v$, $D$ and
$r^2_{\mathrm{sat}}$ on the system size and 
disorder strength, we found the system evolution for a range of values of $N$ and
$\sigma$, and fitted the numerical solutions in the respective time intervals with 
the appropriate power-law  functions of time according to 
Eq.~(\ref{eq:ballistic}), Eq.~(\ref{eq:diffusive}), and Eq.~(\ref{eq:saturation}). The
dependence of all the dynamical characteristics on the two system parameters turns out to
be a power law over at least a decade of parameter variation in each case, with the
power-law exponent very close to an integer or a simple fraction. 

\begin{figure}[tb]
\begin{center}
    \includegraphics[width=\columnwidth]{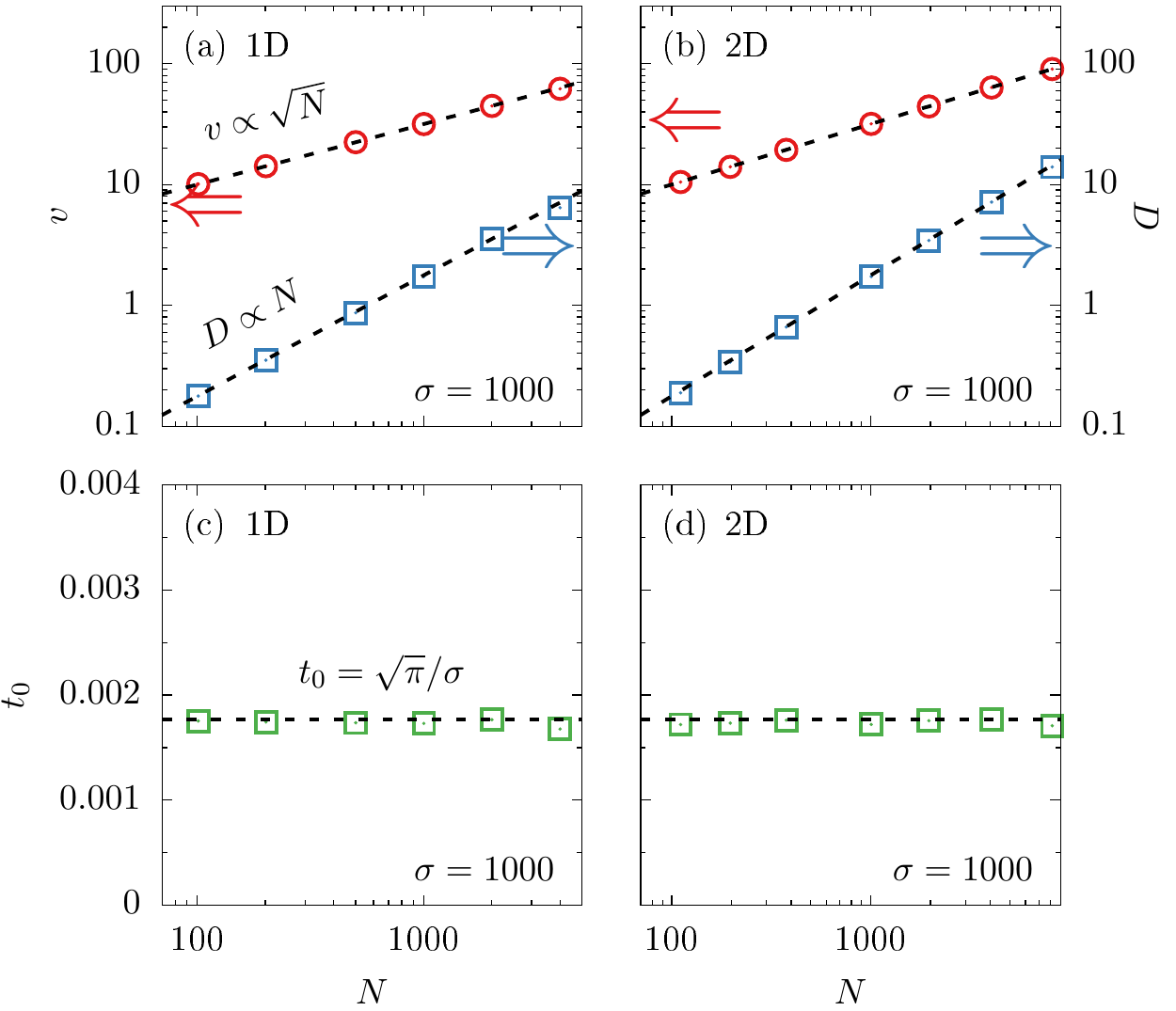}
\end{center}
    \caption{Size dependence of the dynamical coefficients. (a,b) The ballistic velocity
      (green circles, right axis) and the diffusion coefficient (red squares, left axis) as
    a function of the number of sites. (c,d) The ballistic-to-diffusive crossover time as
    a function of the number of sites. Lines show the analytical results from the
    ,,central atom model'' (Sec.~\ref{sec:regimes}).}
   \label{fig:veloctity_diffusion_coefficient}
\end{figure}

The size dependence of the velocity and diffusion coefficient is depicted in
Fig.~\ref{fig:veloctity_diffusion_coefficient}(a,b). 
The velocity increases with the number of atoms like $N^{1/2}$, while the diffusion
coefficient grows linearly with the number of sites both in 1D and 2D. 
As a consequence [Eq.~(\ref{eq:t0})], the first cross-over time is size independent, see 
Fig.~\ref{fig:veloctity_diffusion_coefficient}(c,d).

\begin{figure}[tb]
\begin{center}
    \includegraphics[width=\columnwidth]{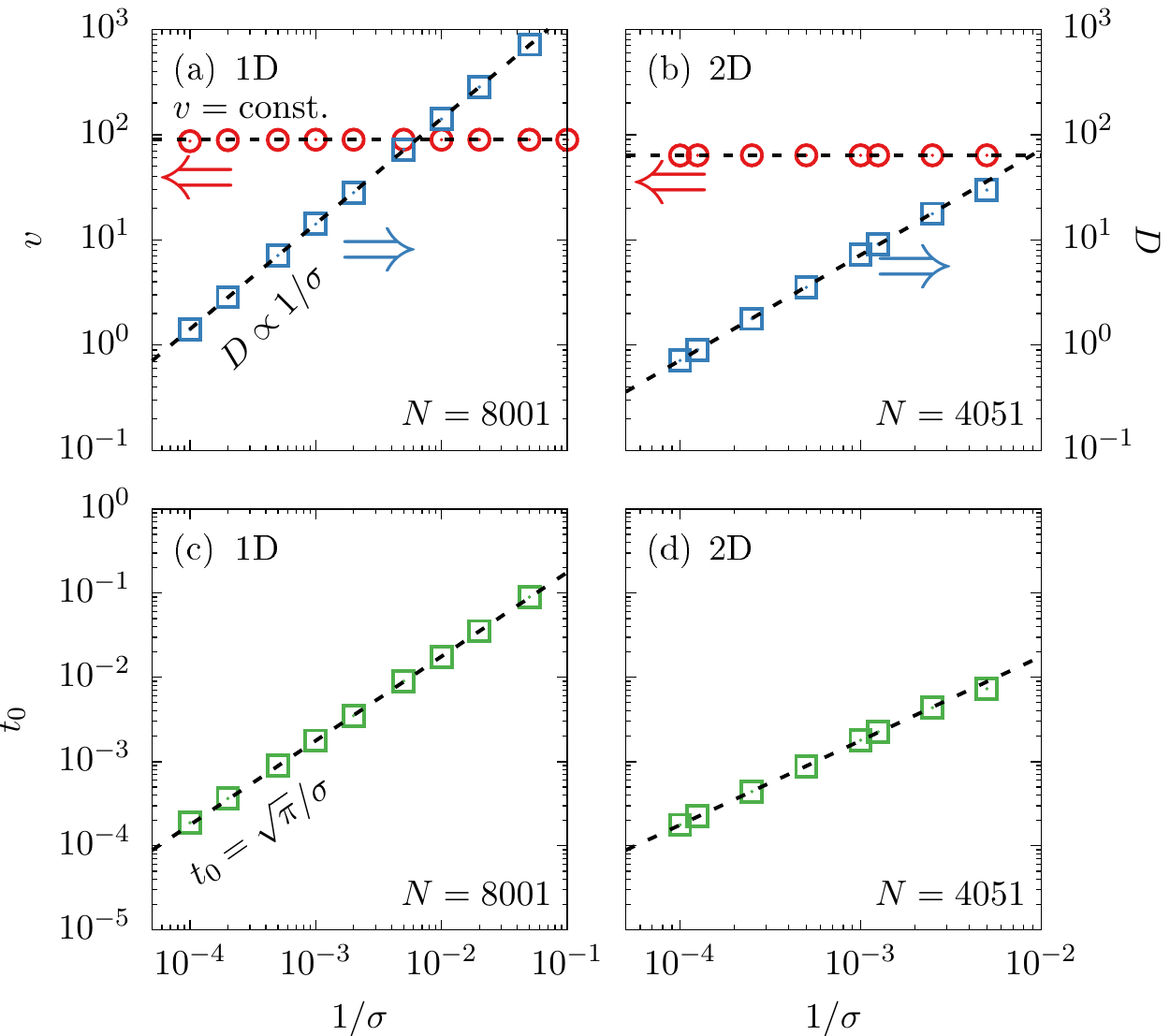}
\end{center}
    \caption{Dependence of the dynamical coefficients on the disorder strength. (a,b) The
      ballistic velocity 
      (red circles, left axis) and the diffusion coefficient (blue squares, right axis) as
    a function of the standard deviation of the on-site energies. (c,d) The
    ballistic-to-diffusive crossover time as 
    a function of the standard deviation of the on-site energies.  Lines show the analytical results from the
    ,,central atom model''.} 
   \label{fig:veloctity_diffusion_coefficient_sigma}
\end{figure}

The dependence of the dynamical parameters on the disorder strength is shown in
Fig.~\ref{fig:veloctity_diffusion_coefficient_sigma}(a,b). 
While the velocity remains $\sigma$-independent, as already concluded above, the diffusion
coefficient is inversely proportional to $\sigma$ for any system dimension. 

\begin{figure}[tb]
    \centering
    \includegraphics[width=\columnwidth]{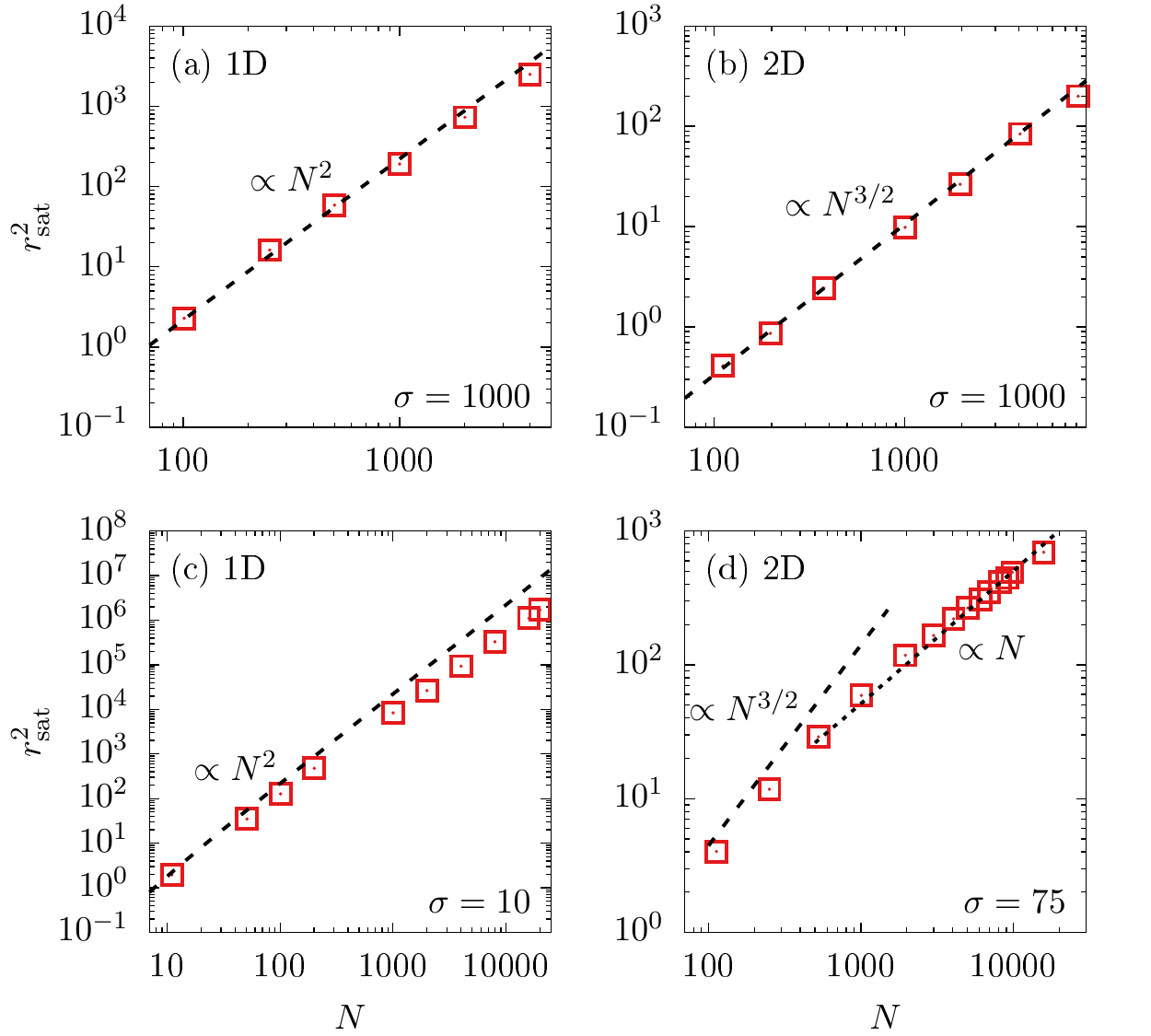}
    \caption{The dependence of the diffusion range on the number of sites in one (a,c) and
      two (b,d) dimensions.  Dashed lines show the analytic results from the
    ,,central atom model''. Dotted line in (d) is the linear fit to to the further part of the plot.}
   \label{fig:saturation_size}
\end{figure}

\begin{figure}[tb]
    \centering
    \includegraphics[width=\columnwidth]{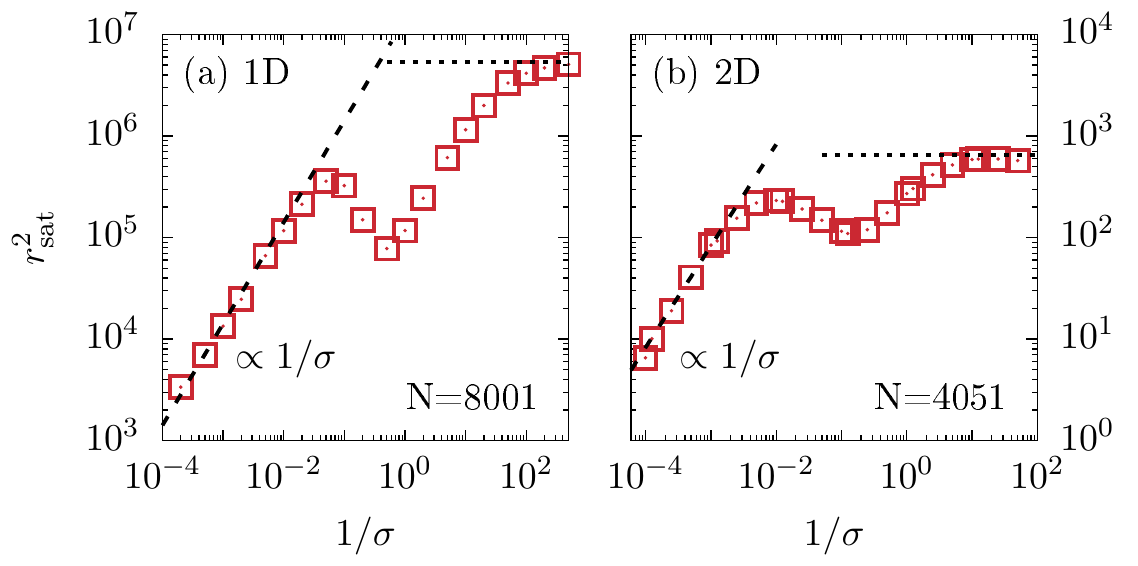}
    \caption{The dependence of the diffusion range on the standard deviation of the
      on-site energies in one (a) and two (b) dimensions.  Solid lines show the analytical results from the
    ,,central atom model''. Dotted lines represent the values corresponding to uniform
    distribution over the system.}
  \label{fig:saturation_sigma}
\end{figure}
 
Finally, in Fig.~\ref{fig:saturation_size} and Fig.~\ref{fig:saturation_sigma} we analyze
the dependence of the saturation level $r^2_{\mathrm{sat}}$ on the system size and
disorder strength, respectively. In 1D, the value of $r^2_{\mathrm{sat}}$ 
grows as $N^{2}$ for large disorder (Fig.~\ref{fig:saturation_size}(a)), while for
moderate disorder one observes an 
approximately power-law dependence with a lower exponent
(Fig.~\ref{fig:saturation_size}(c)). In 2D the dependence is  
$r^2_{\mathrm{sat}} \propto N^{3/2}$ for strongly disordered systems
(Fig.~\ref{fig:saturation_size}(b)) and for sufficiently short chains at weaker disorder
(Fig.~\ref{fig:saturation_size}(d)). This kind of power-law dependence in
two dimensions means that the saturation level grows faster than the system size, hence
the excitation must reach the border of the system for sufficiently large number of sites (on
the order of $\sigma^2$). 
From that point saturation level starts to grow linearly with the system size as it is
clear from Fig.~\ref{fig:saturation_size}(d). In other words, the saturation level is
bounded by the system size. 
Both in 1D and 2D, the value of $r^2_{\mathrm{sat}}$  turns out to
be proportional to $1/\sigma$ as long as $\sigma\gg 1$, while it starts to oscillate and
reaches a constant value as $\sigma\to 0$. The latter property simply reflects the uniform 
spreading of the excitation across the system, characteristic of an unperturbed lattice,
which roughly sets the upper 
limit on the mean-square displacement. The values corresponding to the uniform
distribution are $R^{2}/3$ and $R^{2}/2$ for $d=1$ and $d=2$, respectively, and are marked
with horizontal dotted lines in Fig.~\ref{fig:saturation_sigma}. 
The crossover time $t_{1}$ is then proportional to $N$ and $N^{1/2}$ in 1D and 2D
respectively, that is, to the linear size 
of the system in both cases, and is independent of disorder, as follows from Eq.~(\ref{eq:t1}). 

\begin{figure}[tb]
    \centering
    \includegraphics[width=\columnwidth]{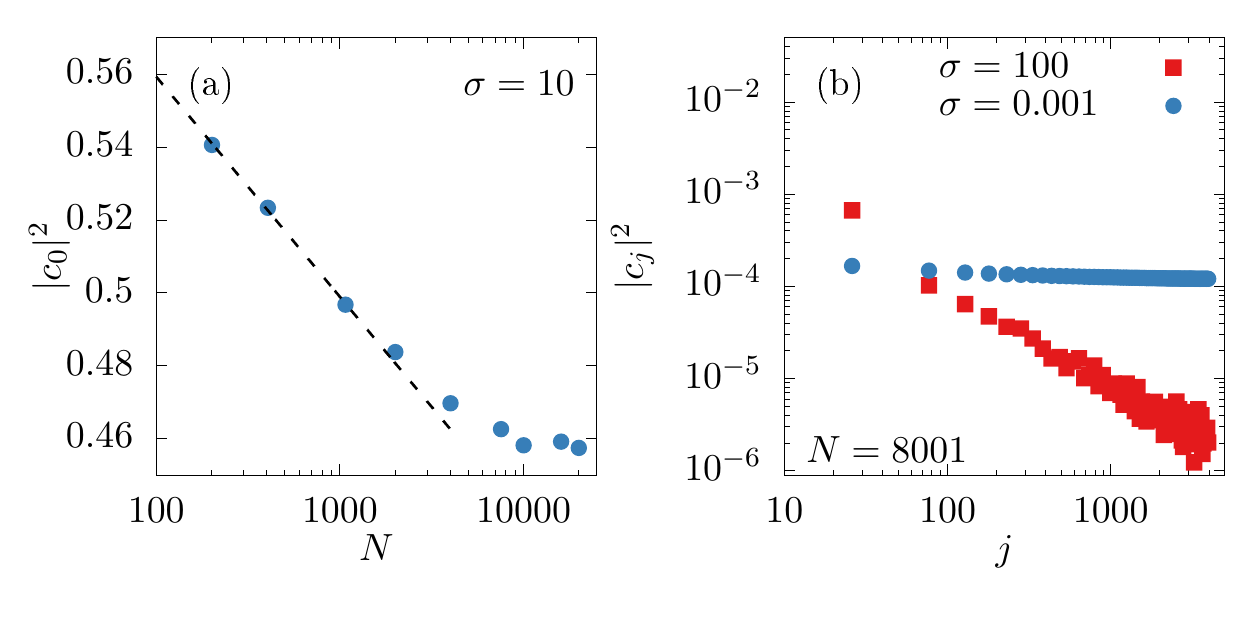}
    \caption{(a) The final (long-time asymptotic) occupation of the initial site in a
      one-dimensional chain as a function of the chain length for $\sigma=10$. (b) The
      dependence of the asymptotic occupation on the site 
      index in a one-dimensional chain for the values of $N$ and $\sigma$ as shown. The
      dashed line shows a logarithmic trend.}
  \label{fig:distrib-1D}
\end{figure}

By a simple lowest-order ``resonance counting'' argument, the number of sites 
resonant to the central (initially occupied) one at a distance $x$ is proportional to $1/x$. 
For a 1D system, as the chain gets longer, due to occupation spreading among these resonant sites, the
occupation of the distant sites would then grow as $\ln N$, the value of $r^2_{\mathrm{sat}}$
would grow as $N^{2}$, and the long-time occupations would be distributed as $1/x$. This
prediction for $r^2_{\mathrm{sat}}$
agrees with the simulation results for very large disorder but discrepancy is visible
at $\sigma=10$ (Fig.~\ref{fig:saturation_size}(c)). The growing occupation of the distant
sites should suppress the final occupation of the initial site $|c_{0}|^{2}$ (the survival
probability) as $1-A\ln N$, until the survival probability is reduced enough for the lowest
order approach to break down. This is indeed confirmed by simulation results
presented in 
Fig.~\ref{fig:distrib-1D}(a), where the logarithmic dependence is represented by the
dashed trend line, although  
the narrow range of variability of $|c_{0}|^{2}$ may be insufficient to rigorously prove the subtle
logarithmic dependence. The saturation at $N\gtrsim 10^{4}$ appears at a rather high value
of the survival probability. The distribution of occupations at saturation (asymptotic
long-time limit) indeed shows a distinct power-law
character over many orders of magnitude of the chain length, with an exponent very close
to 1 for strong disorder (red squares in Fig.~\ref{fig:distrib-1D}(b)). This behavior is
characteristic 
of the strong disorder regime, corresponding to the power-law dependence of
$r^2_{\mathrm{sat}}$ on $\sigma$ (left asymptotics of Fig.~\ref{fig:saturation_sigma}(a)). It
should be contrasted with the low-disorder limit (right asymptotics in
Fig.~\ref{fig:saturation_sigma}(a)), where the occupations are equally spread all over the
chain, as we have already inferred from the asymptotic value  (blue circles in
Fig.~\ref{fig:distrib-1D}(b)).

\begin{figure}[tb]
    \centering
    \includegraphics[width=\columnwidth]{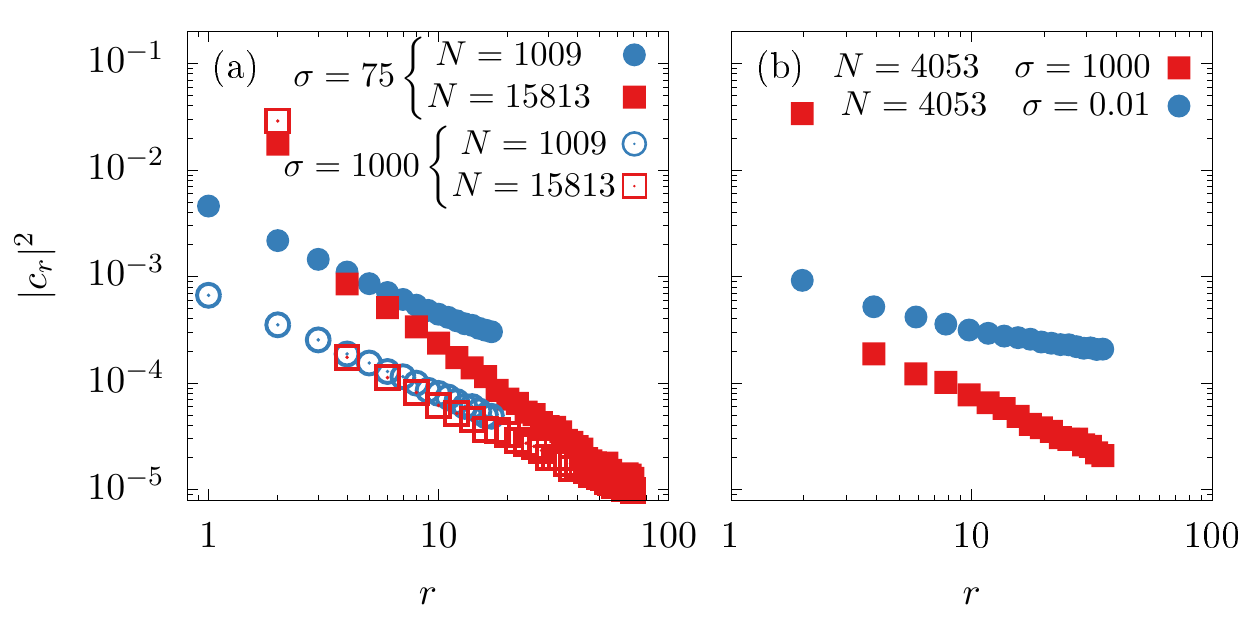}
    \caption{The dependence of the asymptotic occupation on the distance from the
      initially occupied site in a 2-dimensional system: (a) for a fixed $\sigma$  and two
      values of $N$ as shown; (b) for a fixed $N$  and two values of $\sigma$ as shown.} 
  \label{fig:distrib-2D}
\end{figure}

A similar power-law localization of the asymptotic state is observed
in a 2D system, as shown in Fig.~\ref{fig:distrib-2D}, where we plot
the average occupation of a site as a function of its distance from
the central site. As can be seen in Fig.~\ref{fig:distrib-2D}(a) (full symbols), for
a moderate disorder strength, the exponent of this power-law dependence
increases by magnitude as the system size grows. The values from fitting to the
power-law part of the data obtained from simulations range from $-1.0$ for $N=197$ to $-1.7$ for
$N=32017$ (fitting was performed using the central part of the data points, showing the
clear power-law dependence and may be slightly affected by the choice of the cut-offs). All these
values are above $-2$ and therefore correspond  
to heavy tail distributions in two 
dimensions that would have a divergent norm if extrapolated to infinite size. The values
seem to be roughly proportional to $\ln N$ over the range of system sizes available for
numerical simulations and we were not
able to reliably determine the asymptotic value of the exponent as $N\to \infty$. 

A different situation is observed for strongly disordered systems (empty symbols in
Fig.~\ref{fig:distrib-2D}(a)). Here the slope of the power-law dependence is apparently
constant and indeed, the exponents obtained from fitting oscillate (due to inherent
randomness and fitting uncertainty) very close to the value of $-1$, which precisely
corresponds to the $r^{2}_{\mathrm{sat}}\propto N^{3/2}$ dependence in 
Fig.~\ref{fig:saturation_size}(b). This means that in this range of system sizes, the
asymptotic diffusion range grows with the system size by extending the $\propto 1/r$
dependence of occupation to larger and larger distances at the cost of the central site
occupation, until the $\propto N^{3/2}$ dependence breaks down, similar to the situation
in Fig.~\ref{fig:saturation_size}(d) but far beyond the range of system sizes accessible
in simulations. When comparing the power-law dependence at the two disorder strengths one
arrives at the somewhat
unexpected conclusion that, from the formal point of view focused merely on the power-law
exponent, localization in moderately disordered systems is stronger (the absolute value of
the exponent is larger) than in heavily
disordered ones. This is obviously not true in terms of the actual values of the
occupations that decrease as the disorder grows, which simply means that the survival
probability at the initial site grows with disorder, as expected. Interestingly, the
occupations of the most remote sites are similar for the two different disorder regimes
shown in Fig.~\ref{fig:distrib-2D}(a).

The trend in the dependence of the average asymptotic distribution of occupations on the
disorder strength demonstrated above cannot remain valid towards weaker disorder
strengths, as the occupation should spread across all the system in the limit of
unperturbed chain. Indeed, as shown in Fig.~\ref{fig:distrib-2D}(b), for a very weak
disorder, the system tends to a uniform distribution but, 
rather surprisingly, even for $\sigma=0.01$ a weak localization effect is still visible. 

In addition to the simulations of the full model discussed so far, we have also studied
the 
dynamics of a ``central atom model'' in which the central (initially excited) site is
coupled to all the other sites in the system as in the full model, but the other sites are
not coupled with one another, i.e., $V_{\alpha\beta} = 0$ if $\alpha\neq 0$ and $\beta\neq
0$. The results are shown in Fig.~\ref{fig:average_square_distance} with dotted lines. One
can see that the system evolution is nearly the same in both models in this parameter
range, which means that the
dynamics in the strongly disordered case is dominated by direct jumps to remote places,
which is possible due to the long-range coupling. 

In the next section we show that the dynamical parameters can be related to the system size
and disorder strength and the analytical relation between the power-law exponents and the
system dimension can be found as long as the ``central atom model'' is valid. 

\section{Approximate analytical solution}\label{sec:analytic}

In this section we present an approximate analytic approach to the considered problem,
which becomes possible within the simplified ``central atom model'' in the limit of strong
disorder.

\subsection{Solution of the Equation of Motion}
\label{sec:solution}

The equation of motion for the coefficients of the expansion defined in
Eq.~(\ref{Psi-expansion}) has the form
\begin{equation}
    i\dot{c}_\alpha(t) = \epsilon_\alpha c_{\alpha}(t) +  \sum\limits_{\beta} V_{\alpha\beta} c_\beta(t), \qquad c_\alpha(0) = \delta_{\alpha0}.\label{eq:state_eq}
\end{equation}

We define the Laplace transform $f_\alpha(s)$ of an amplitude $c_\alpha(t)$,
\begin{equation}
f_{\alpha}(s) = \int\limits_0^{\infty} e^{-st} c_{\alpha}(t) \mathrm{d}t,
\end{equation}
where $s$ is a complex variable.
Equation \eqref{eq:state_eq} in terms of the Laplace transform is
\begin{equation}
    f_\alpha(s) = \dfrac{i\delta_{0\alpha}}{is-\epsilon_\alpha} + \sum\limits_{\beta\neq \alpha}V_{\alpha\beta}f_\beta(s).
    \label{eq:state_equation_in_Laplace_transform}
\end{equation}
Eq.~\eqref{eq:state_equation_in_Laplace_transform} can be iteratively expanded in series
depending only on the central-site term $f_0$, 
\begin{align}
    f_\alpha(s) &= \dfrac{V_{\alpha0}}{is-\epsilon_\alpha}f_0(s) \label{eq:series_f_alpha}\\ 
&\quad + \sum\limits_{\beta\neq \alpha,\beta\neq
  0}\dfrac{V_{\alpha\beta}}{is-\epsilon_\beta}\dfrac{V_{\beta0}}{is-\epsilon_\beta}f_0(s)+\ldots,\quad
  \alpha\neq 0 \nonumber
\end{align}
and 
\begin{align}
    f_0(s) &= \dfrac{1}{is-\epsilon_0} + \left[\sum\limits_{\beta\neq 0}
             \dfrac{V_{0\beta}V_{\beta0}}{(is-\epsilon_0)(is-\epsilon_\beta)}\right. \label{eq:series_f_0}\\
  &\quad + \left.\sum\limits_{\gamma\neq \beta, \gamma\neq
  0}\dfrac{V_{0\beta}V_{\beta\gamma}V_{\gamma0}}{(is-\epsilon_0)(is-\epsilon_\beta)(is-\epsilon_\gamma)}
+\dots\right]f_0(s)  \nonumber.
\end{align}
The series have the number of terms of the order of $N^{N}$ and the resulting system of
equations cannot be solved in a simple manner. 
The problem simplifies considerably in the ``central atom approximation''.
Then, only the first term in Eq.~\eqref{eq:series_f_alpha} remains, while in
Eq.~\eqref{eq:series_f_0} only the first term and the first sum survive. 
After these simplification one can write $f_\alpha(s)$ as
\begin{gather}
    f_\alpha(s) = \dfrac{iV_{\beta0} \prod\limits_{\beta\neq 0, \beta \neq \alpha}\left(is-\epsilon_\beta\right)}{\prod\limits_\beta \left(is-\epsilon_\beta\right)-\sum\limits_{\beta\neq 0}\prod\limits_{\gamma\neq \beta,\gamma\neq 0}V_{0\beta}V_{\beta0}\left(is - \epsilon_\gamma\right)}.
        \label{eq:fj_final}
\end{gather}

Next, we transform back to the time domain, by means of the Mellin's formula
\begin{gather}
    c_\alpha(t) = \frac{1}{2\pi i}\lim_{T\to\infty}\int\limits_{\gamma-iT}^{\gamma+iT}e^{st}f_\alpha(s)\mathrm{d}s.
\end{gather}
To perform the integral we employ residue theorem and obtain
\begin{gather}
	c_\alpha(t) = \sum\limits_n b_{n}^{(\alpha)}e^{-i z_n t},
\end{gather}
where $z_n = is_n$ and $s_n$ are the poles of the analytic function given by Eq.~\eqref{eq:fj_final}. 
All $z_n$ must be real.
In addition,
\begin{gather}
	b_n^{(\alpha)} = 2\pi i\mathrm{Res}_{z_n} f_\alpha(z) = \dfrac{V_{\alpha0}\prod\limits_{\beta\neq 0, \beta\neq \alpha}(z_n-\epsilon_\beta)}{\prod\limits_{\beta\neq n}(z_n-z_\beta)},
\end{gather}
where $\mathrm{Res}_{z_n} f_\alpha(z)$ denotes the residue of $f_\alpha(z)$ at $z_n$.

As long as the coupling is small in comparison to $\sigma/N$, the roots of the denominator
of Eq.~\eqref{eq:fj_final} lie in close vicinity to bare energies $\epsilon_\alpha$, as
compared to the typical distance between these roots. 
Hence, one can associate each pole with the nearest bare energy, align the numbering and
assume $\dfrac{z_n-\epsilon_\beta}{z_n-z_\beta} \approx 1$, whenever $n\neq\beta$. With
this approximation one finds 
\begin{equation*}
b_{n}^{(\alpha)}\approx 
\left\{\begin{array}{cl}
\dfrac{V_{\alpha0}}{z_0-z_\alpha},& \textrm{$n=0, n\neq \alpha$},\\[0.4cm]
\dfrac{V_{\alpha0}}{z_n-z_0},& \textrm{$n=\alpha\neq 0$},\\
0,& \textrm{otherwise.}
\end{array}\right.
\end{equation*}
The amplitudes $c_\alpha(t)$ then take the form
\begin{align*}
	c_\alpha(t) &= b_0^{(\alpha)} e^{-iz_0 t}+b_\alpha^{(\alpha)}e^{-iz_\alpha t}\\
&= e^{-i z_0 t} \frac{V_{\alpha 0}}{z_0-z_\alpha}\left(1-e^{-i(z_0-z_\alpha)t}\right)
\end{align*}
and the occupation of the site $\alpha$ becomes
\begin{gather}
	|c_\alpha(t)|^2 = |V_{\alpha0}|^2\frac{\sin^2\left(\delta z_\alpha t/2\right)}{\left(\delta z_\alpha/2\right)^2},
\end{gather}
where $\delta z_\alpha = z_0-z_\alpha$.
Upon inserting this result to the Eq.~\eqref{eq:r2_aver} we obtain
\begin{gather}
	\left\langle r^2 (t) \right\rangle = 
% 	\left\langle \sum\limits_{\alpha} r_{\alpha}^2 \left|c_\alpha(t)\right|^2\right\rangle \nonumber\\
	 \left\langle \sum\limits_r r^2 \left|V_r\right|^2 \sum\limits_{k=1}^{n_r} \dfrac{\sin^2(\delta z_{k}(r)t/2)}{\left(\delta z_{k}(r)/2\right)^2} \right\rangle,
\end{gather}
where we decomposed the sum over all the sites into subsets of $n_r$ sites lying at a
distance $r$ from the origin. 
All the sites at a given distance have the same coupling to the central dot $V_r=1/r$.
$\delta z_k(r)$ are the values of $\delta z_k$ for sites lying at distance $r$ from the origin.
$\delta z(r)$ can be considered random variables with a certain probability density $f_r(\delta z)$.
In the continuum approximation one then obtains
\begin{gather}
\langle r^2 (t) \rangle =
\int\limits_0^R \zeta_d d r^{d-1} \mathrm{d}r 
\int\limits_{-\infty}^{\infty} f_r(u) \dfrac{\sin^2(u t/2)}{(u/2)^2}\mathrm{d}u,
\label{eq:intergral_explaining_results}
\end{gather}
where $\zeta_d r^d$ is the number of sites lying inside a $d\textrm{-dimensional}$ sphere.

At large $\delta z$, the poles are shifted negligibly from the bare energies, hence the
distribution function is close to the on-site energy difference 
distribution $f_{\infty}(\delta\epsilon)$ (here $\infty$ refers to infinite distance,
hence vanishing coupling, and $\delta\epsilon$ is the bare energy difference), which is a
Gaussian distribution with  
the standard deviation $\sqrt{2}\sigma$. However, at $\delta z\sim V_r$ the pole
probability distribution must reflect the level repulsion. Its form can be 
found by noting that a pair of sites with a bare energy difference $\delta\epsilon> 0$
coupled with a coupling strength $V$ gives rise to a pair of poles separated by $\delta z
= \sqrt{(\delta\epsilon)^2 + 4V^2}$. From this, the cumulative distribution function for
$\delta z$ follows in the form 
\begin{gather}
    P(0<\delta z_k(r)<u) = \!\!\int\limits_0^{\sqrt{u^2-4V_r^2}} \!\!
 f_\infty(x) \mathrm{d}x,\quad \abs{u}>2V_r,
\end{gather}
and the corresponding probability density is
\begin{gather}
    f_r (u) = \left\{\begin{array}{lc}
         f_\infty(\sqrt{u^2-4V_r^2}) \frac{|u|}{\sqrt{u^2-4V_r^2}}, & \abs{u} > 2V_r,\\
         0, & \abs{u} \leq 2V_r.
    \end{array}\right.
    \label{eq:distribution}
\end{gather}
The above distribution corresponds to the normal distribution of standard deviation
$\sigma_u = \sqrt{2}\sigma$ having a gap around zero of width $4V_r$. 
The total distribution for a system of radius $R$ including all possible distances $r$ has
a gap of width $4V_R$. 

\subsection{Regimes of propagation}
\label{sec:regimes}

Eqs.~\eqref{eq:intergral_explaining_results} and \eqref{eq:distribution} allow us to
explain the three time phases in the excitation evolution. 

\begin{figure}[tb]
\begin{center}
    \includegraphics[width=\columnwidth]{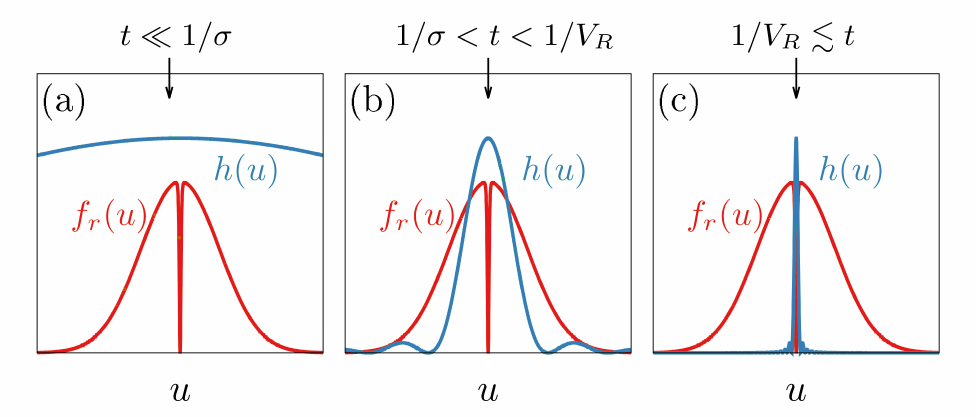}
\end{center}
    \caption{Schematic plots of the probability density $f_r$ (red, the same for three
      cases) and the function $h(u)$ (blue) corresponding to three phases of propagation:
      (a) ballistic, (b) diffusive, (c) saturation.} 
    \label{fig:rysunek_rozklady}
\end{figure} 

For very short times, $t\lesssim 1/\sigma$, the function 
$h(u)=\sin^2(ut/2)/(u/2)^{2}$ is slowly varying in $u$ and can be approximated by
$h(u)\approx t^{2}$ over the whole width of the distribution $f_{r}(u)$, as schematically
shown in Fig.~\ref{fig:rysunek_rozklady}(a). Eq.~(\ref{eq:intergral_explaining_results}) then immediately
yields $\langle r^2 (t) \rangle = v^{2}t^{2}$, that is, ballistic propagation with the
constant velocity
\begin{equation}
    v^2 = \int\limits_0^{R} \zeta_d d r^{d-1} \mathrm{d}r \int\limits f_\infty(u)
    \mathrm{d}u = \zeta_d R^d  =  N. 
\label{eq:velocity_squared}
\end{equation}
This dependence is shown as dashed lines in
Fig.~\ref{fig:veloctity_diffusion_coefficient}(a,b), which perfectly follows the
numerical data.

In the intermediate time range, $1/\sigma < t < 1/V_{R}$, the function $h(u)$ probes the
central part of the distribution but is still insensitive to the narrow central gap
(Fig.~\ref{fig:rysunek_rozklady}(b)). Then 
the integral over $u$ in Eq.~(\ref{eq:intergral_explaining_results}) can be approximated
by
\begin{displaymath}
\int\limits_{-\infty}^{\infty} f_r(u) \dfrac{\sin^2(u t/2)}{(u/2)^2}\mathrm{d}u \approx
\int\limits_{-\infty}^{\infty} f_\infty(u) 2\pi \delta(u)  \mathrm{d}u = 2\pi t f_{\infty}(0),
\end{displaymath}
from which one finds the diffusive transport $\langle r^2 (t) \rangle =Dt$, with 
\begin{equation} \label{eq:diffusion_coefficient}
  D = \int\limits_0^{R} \mathrm{d}r \zeta_d d r^{d-1}2\pi t f_{\infty}(0) 
    = \frac{\sqrt{\pi} N}{\sigma}.
\end{equation}
Again, this dependence on $N$ and $\sigma$ is shown as dashed lines in
Fig.~\ref{fig:veloctity_diffusion_coefficient}(a,b) and in
Fig.~\ref{fig:veloctity_diffusion_coefficient_sigma}(a,b). The agreement with the numerical data is excellent.
%except for the smallest values of $\sigma$, where the
%analytical result overestimates the diffusion coefficient. 

Using Eq.~\eqref{eq:t0}, the crossover time between ballistic and
diffusive phases is obtained as 
\begin{gather}
    t_0 = \sqrt{\pi}/\sigma,
\end{gather}
which agrees with our simulation results, as shown in
Fig.~\ref{fig:veloctity_diffusion_coefficient_sigma}(b), for sufficiently large values of
$\sigma$. The range of the ballistic transport is therefore 
$\langle r^2 (t_{0}) \rangle = v^{2}t_{0}^{2}=N\sqrt{\pi}/\sigma$. In 1D it is always much
smaller than the system size if the disorder is strong.

Finally, at $t\approx 1/V_{R}$, the function $h(u)$ becomes as narrow as the the gap in
the density function, hence its central part does not contribute, while its oscillating
tails are averaged to  $\tilde{h}(u) = (1/2)/u^2$  (see
Fig.~\ref{fig:rysunek_rozklady}). There is no time dependence in this limit, 
which results in the saturation of $\langle r^2 \rangle$.  
The saturation level is then estimated from Eq.~\eqref{eq:intergral_explaining_results} as 
$2\int_{V_r}^{\infty}f_r(u) \tilde{h}(u) \mathrm{d}u =
\frac{\sqrt{\pi}r}{2\sigma }\exp[1/(\sigma^2r^2)]\mathrm{erfc}[1/(\sigma r)]$
The resulting saturation saturation value is
\begin{align}
    r^2_\mathrm{sat} &= \langle r^2 (t) \rangle =
\int\limits_0^R \zeta_d d r^{d}  \frac{\sqrt{\pi}}{2\sigma} \exp\left(\frac{1}{\sigma^2r^2}\right)\mathrm{erfc}\left(\frac{1}{\sigma r}\right) \mathrm{d}r \nonumber \\ &\approx  \frac{\sqrt{\pi}\zeta_d}{2\sigma} \frac{d}{d+1}R^{d+1} 
% - \frac{\zeta_d}{\sigma^2}R^d
%\frac{\zeta_d R^{d+1}
 % }{\sigma\sqrt{\pi}}\left\{\exp\left[-\left(\frac{1}{R\sigma}\right)^2\right]\right. \left.-
%  \frac{\sqrt{\pi}}{R\sigma}\mathrm{erfc}\left(\frac{1}{R\sigma}\right)\right\} \nonumber \\
%&\approx \frac{\zeta R^{d+1}}{\sqrt{\pi}\sigma}, 
\label{eq:saturation_approx}
\end{align}
where we took into account that $V_R/\sigma =1/R\sigma \ll 1$ in the high disorder energy
regime, so the last two terms in the integral tends to unity (in the zeroth order of Taylor expansion). 
% was expanded into series up to the first order of $1/\sigma r$.

The dependence from Eq.~\eqref{eq:saturation_approx} is marked by dashed lines in
Fig.~\ref{fig:saturation_size} and Fig.~\ref{fig:saturation_sigma} and agrees very well
with the simulation result as long as the size is not too large and the disorder is
strong. 

The onset of saturation is determined by the continuity of $\langle r^2 \rangle$,  $Dt_1 =
r^2_\mathrm{sat}$.
Combining Eq.~\eqref{eq:saturation_approx} with \eqref{eq:diffusion_coefficient} one obtains
\begin{equation}
    t_1 = \dfrac{1}{\pi V_R}, \label{eq:t_1_from_analytic}
\end{equation}
in agreement with the simulations.

Within the ``central atom model'', the 
asymptotic occupation of a site at a distance $r$ from the origin is
\begin{align}
  \langle |c_r|^2 \rangle_\mathrm{sat} & =
\dfrac{2}{r^2}\int\limits_{2V_r}^\infty f_r(u)\dfrac{2}{u^2}\mathrm{d}u \nonumber\\
& =\dfrac{\sqrt{\pi}}{2\sigma r}
\exp\left(\dfrac{1}{\sigma^2r^2}\right)\mathrm{erfc}\left(\dfrac{1}{\sigma r}\right)
\approx \dfrac{\sqrt{\pi}}{2\sigma r}.
\end{align}
This $\propto 1/r$ trend obtained from our approximate analytical solution agrees with the
numerical data in 1D, shown in Fig.~\ref{fig:distrib-1D} and 
corresponds to the survival probability $1-A\ln N$ in the strong disorder regime, which
is consistent with the simulation data in Fig.~\ref{fig:distrib-1D}(a). In 2D, as we have
seen in Fig.~\ref{fig:distrib-2D}, the same dependence is obtained for very strongly
disordered systems.

\subsection{Discussion}
\label{sec:discussion}

As we have seen, the analytical formulas agree very well with the simulation results only within a certain limits of system size and disorder strength. One discrepancy appears when the disorder becomes weak (see Fig.~\ref{fig:saturation_sigma}). 
This is obvious, as our ``central atom model''  is essentially based on the assumption that coupling is a perturbation to the on-site energies, which requires a strong disorder. 
The second discrepancy appears for long chains in Fig.~\ref{fig:saturation_size}.
For $d\ge 2$, $r^2_\mathrm{sat} \sim R^{d+1}$, Eq.~\eqref{eq:saturation_approx}
predicts that the asymptotic diffusion range grows faster than the system size.
This cannot be true for an arbitrary system size and, indeed, the trend in simulations
changes at a certain system size Fig.~\ref{fig:saturation_size}(d), which is not captured
by the ``central atom'' approximation.
We note that the limit of validity
of our approximation is $\sigma/N \sim V_{R} = 1/R$ or, using $N\sim R^{d}$,
$R^{d-1}\sim\sigma$. 
At this limit,
% Eq.~(\ref{eq:saturation_approx}) yields 
$\langle r_{\mathrm{sat}}^{2} \rangle\sim R^{2}$, which assures consistency of our conclusions.  
On the other hand, for a one-dimensional chain, $r^2_\mathrm{sat}\sim R^2$, hence
the asymptotic range grows linearly with the system size. 
Eq.~\eqref{eq:saturation_approx} is valid for $\sigma\gg1$, hence $r^2_\mathrm{sat}\ll R^{2}$ and
the excitation is effectively trapped around the original site in the chain.

\section{Conclusions}
\label{sec:conclusions}

We have studied the diffusion of an initially localized excitation in finite lattices with
a strong on-site 
disorder and a long-range coupling (hopping) inversely proportional to the distance. We
have shown that the diffusion in such a system takes place in three dynamical stages:
ballistic transport is followed by normal diffusion and then by saturation. The numerical
findings can be understood with the help of an approximate model which is valid in the
strong disorder regime and allows an analytical solution, which relates the dynamical
properties to the system parameters (size and disorder strength).

We have proposed two complementary descriptions of localization. The first one emerges
from the dynamics and consists in analyzing the asymptotic range of diffusion. The other
one consists in studying the spatial distribution of the average occupations of lattice
sites in the long-time limit and is more directly related to the structure of energy eigenstates
of the disordered system. We have shown, both numerically and analytically, that the
diffusion range grows proportionally to the system size and is always much smaller than
the latter in 1D hence, from this point 
of view, the excitation remains effectively localized around the initial site. In
contrast, in 2D systems, the range of diffusion initially grows faster than the system size as the
latter increases, until at a certain system size the growth slows down so that, in
sufficiently strongly disordered systems the range again remains much smaller than the
system size. While tracing the asymptotic diffusion range provides only a single number
characterizing the degree of localization or spreading of the excitation, inspection of
the average profile of occupations offers a more complete spatial picture of the
localization. We have found out that after a sufficiently long time the occupations
stabilize into a heavy-tailed power-law distribution both in 1D and 2D that not only does
not have a second moment but would even have a divergent norm when extrapolated to
infinite system size. Therefore, even if the excitation remains localized in the sense of
showing the diffusion range much smaller than the system size, it does not have any
intrinsic localization length and the diffusion reaches (on the average) an arbitrary site
of the lattice according to the power-law distribution as a function of the distance. 

\begin{acknowledgments}
The authors are grateful to Marcin Mierzejewski for discussions. 

Calculations have been partially carried out using resources provided by Wroclaw Centre for Networking and Supercomputing (http://wcss.pl), grant No. 203.
\end{acknowledgments}

%\bibliographystyle{apsrev4-2}
%\bibliography{mybib}
% \bibliographystyle{ieeetr}

\end{document}